\documentclass[prl,twocolumn,showpacs]{revtex4}% Physical Review Letters

\usepackage{graphicx}% Include figure files
\usepackage{dcolumn}% Align table columns on decimal point
\usepackage{bm}% bold math

\begin{document}

%\preprint{APS/123-QED}

\title{Mott Transition in the Two-Dimensional Hubbard Model}

\author{Masanori Kohno}
\affiliation{WPI Center for Materials Nanoarchitectonics, 
National Institute for Materials Science, Tsukuba 305-0044, Japan}

\date{\today}

\begin{abstract}
Spectral properties of the two-dimensional Hubbard model near the Mott transition are investigated by using cluster perturbation theory. 
The Mott transition is characterized by freezing of the charge degrees of freedom in a single-particle excitation that leads continuously to the magnetic excitation of the Mott insulator. 
Various anomalous spectral features observed in high-temperature superconductors are explained in a unified manner as properties near the Mott transition. 
\end{abstract}


\maketitle
%{\it introduction.$-$}
The metal-insulator transition due to Coulomb interactions between electrons is called the Mott transition. 
In the insulating phase, the charge excitation has an energy gap whereas the spin excitation is usually gapless \cite{AndersonSW}. 
On the other hand, a metal far away from a Mott insulator exhibits free-electron-like behaviors. 
Thus, the question of how the two limits can be reconciled at the Mott transition is a fundamental puzzle in condensed-matter physics. 
In particular, in relation to the anomalous properties of cuprate high-temperature (high-$T_c$) superconductors \cite{ShenRMP,Graf,DagottoRMP}, 
the Mott transition in a two-dimensional (2D) system has been investigated 
from various viewpoints \cite{DagottoRMP,ImadaRMP,KotliarRMP,PhillipsRMP,Eskes,NagaosaRMP,ImadaCofermion,PreussQP,PreussPG,Bulut,Stanescu,SakaiImada,ScalapinoKink,Furukawa,KohnotJ,DagottoFlatband,CPTPRB}. 
However, its nature remains controversial. 
\par
This Letter shows how the single-particle excitation in a 2D metal changes into the spin-wave mode of the Mott insulator \cite{AndersonSW}, 
through analyses of numerical results on the 2D Hubbard model according to the exactly known properties of a one-dimensional (1D) system \cite{Kohno1DHub}. 
In addition, anomalous spectral features observed in high-$T_c$ cuprates, such as the pseudogap, Fermi arc, flat band, doping-induced states, spinon-like and holon-like branches, 
as well as kink and waterfall in the dispersion relation \cite{ShenRMP,Graf,DagottoRMP}, are explained in a unified manner as properties of the 2D Hubbard model near the Mott transition. 
The 2D Hubbard model is defined by the following Hamiltonian: 
$$
{\cal H}=-t\sum_{\langle i,j\rangle\sigma}\left(c_{i\sigma}^{\dagger}c_{j\sigma}+{\mbox {H.c.}}\right)+U\sum_{i}n_{i\uparrow}n_{i\downarrow}-\mu\sum_{i\sigma}n_{i\sigma}, 
$$
where $c_{i\sigma}$ and $n_{i\sigma}$ denote the annihilation and number operators of an electron with spin $\sigma$ at site $i$, respectively, 
and $\langle i, j\rangle$ means that sites $i$ and $j$ are nearest neighbors on a square lattice. The doping concentration is denoted by $\delta$. 
We consider the spectral function defined as $A({\bm k}, \omega)\equiv-\frac{1}{\pi}\mbox{Im}G({\bm k}, \omega)$, 
where $G({\bm k}, \omega)$ denotes the retarded single-particle Green function at momentum ${\bm k}$ and energy $\omega$ \cite{ImadaRMP} 
of the 2D Hubbard model at zero temperature. 
We employ cluster perturbation theory (CPT) \cite{CPTPRB} to obtain $G({\bm k}, \omega)$, 
using ($4\times4$)-site cluster Green functions calculated by exact diagonalization. 
%An advantage is the larger size of clusters treatable with exact diagonalization than that in cellular dynamical mean-field theory \cite{SakaiImada}. 
%The larger size of clusters treatable with exact diagonalization is an advantage of CPT over cellular dynamical mean-field theory \cite{SakaiImada}. 
The size of clusters treatable with exact diagonalization in CPT is larger than that in cellular dynamical mean-field theory \cite{SakaiImada}. 
We consider the properties of hole-doped systems with $t > 0$ for $ 0\le k_y\le k_x\le \pi$ without loss of generality. 
\par
\begin{widetext}

\begin{figure}
\includegraphics[width=15.4cm]{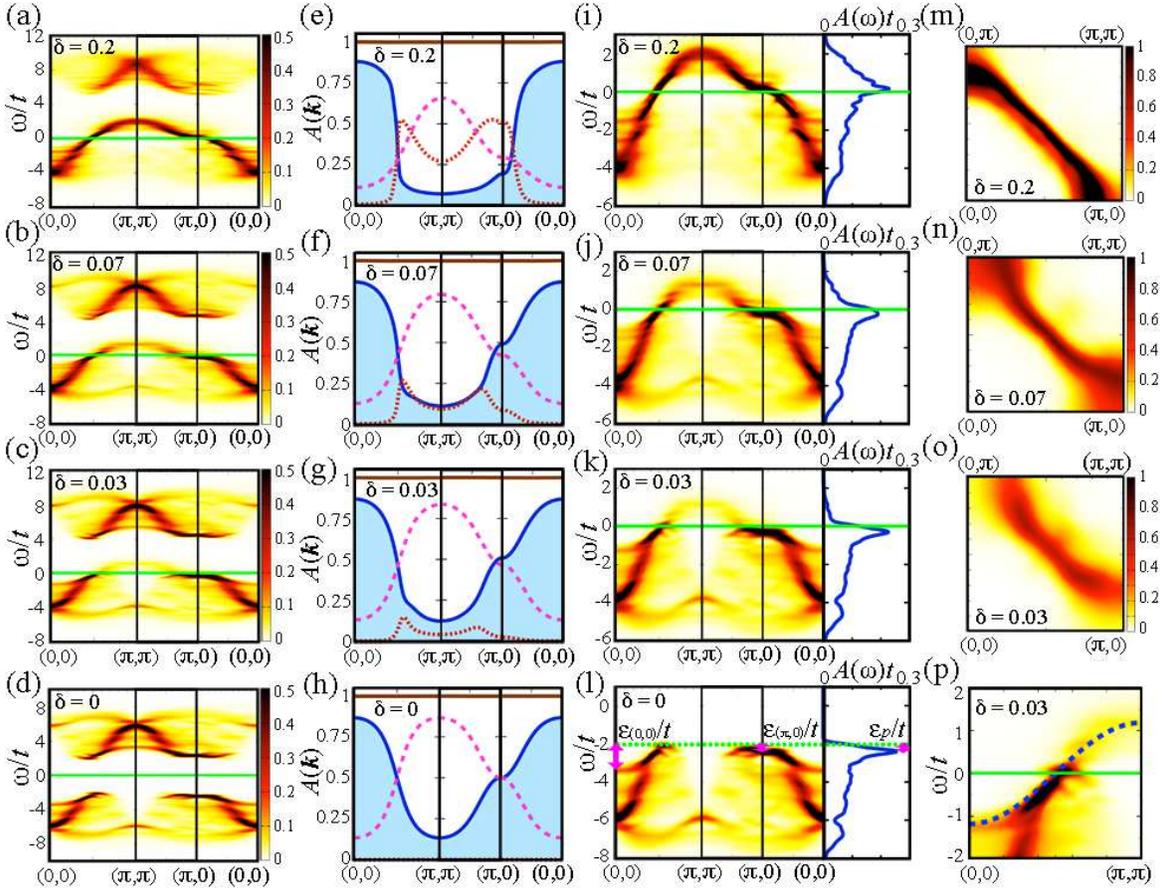}
\caption{Results on the 2D Hubbard model for $U/t = 8$ obtained using cluster perturbation theory. 
(a--d) $A({\bm k}, \omega)t$ and (e--h) $A({\bm k}) [\equiv\int d\omega A({\bm k}, \omega)]$ for $\delta = 0.2$, 0.07, 0.03, and 0 (from above). 
In (e--h), contributions from $\omega < 0$, the lower Hubbard band (LHB) for $\omega > 0$, and the upper Hubbard band are denoted 
by solid blue curves with hatches, dotted red curves, and dashed pink curves, respectively. 
Total spectral weights (solid brown lines) satisfy $A({\bm k}) = 1$ within numerical accuracy. 
(i--l) Close-ups of (a--d) for the LHB. 
Rightmost panels show $A(\omega)t [\equiv \int  d{\bm k}A({\bm k}, \omega)t/(2\pi)^2]$. 
In (l), dotted green line indicates energy of the top of the LHB. Bandwidth of the mode originating from the 1D spinon mode, 
pseudogap defined by the flat mode at ($\pi$,0), and that by the peak in $A(\omega)$ at $\delta=0$ are denoted by $\epsilon_{(0,0)}$, $\epsilon_{(\pi,0)}$ and $\epsilon_p$ 
[pink arrows at (0,0), ($\pi$,0), and in the rightmost panel], respectively. 
%(m--o) $A({\bm k}, \omega \approx 0)t$ for $\delta = 0.2$, 0.07, and 0.03 (from above). 
(m--o) $A({\bm k}, \omega)t$ at $\omega \approx 0$ for $\delta = 0.2$, 0.07, and 0.03 (from above). 
%(p) Close-up of (k) around $\omega \approx 0$ in the (0,0)--($\pi$,$\pi$) direction. 
%(p) Close-up of (k) around $\omega = 0$ in the (0,0)--($\pi$,$\pi$) direction. 
(p) Close-up of (k) near $\omega = 0$ in the (0,0)--($\pi$,$\pi$) direction. 
Dashed blue curve indicates $\epsilon_{2D}(k, k)/t = -\sqrt 2 (v_{2D}/t) \cos k$, where spin-wave velocity of the 2D Heisenberg model $v_{2D} \approx 1.18\sqrt 2 J$ \cite{Singh} ($J = 4t^2/U$). 
Solid green lines show $\omega = 0$. Gaussian broadening is used with standard deviation $\sigma = 0.1t$.}
\label{fig:Fig1}
\end{figure}
\end{widetext}
\begin{figure}
\includegraphics[width=8.4cm]{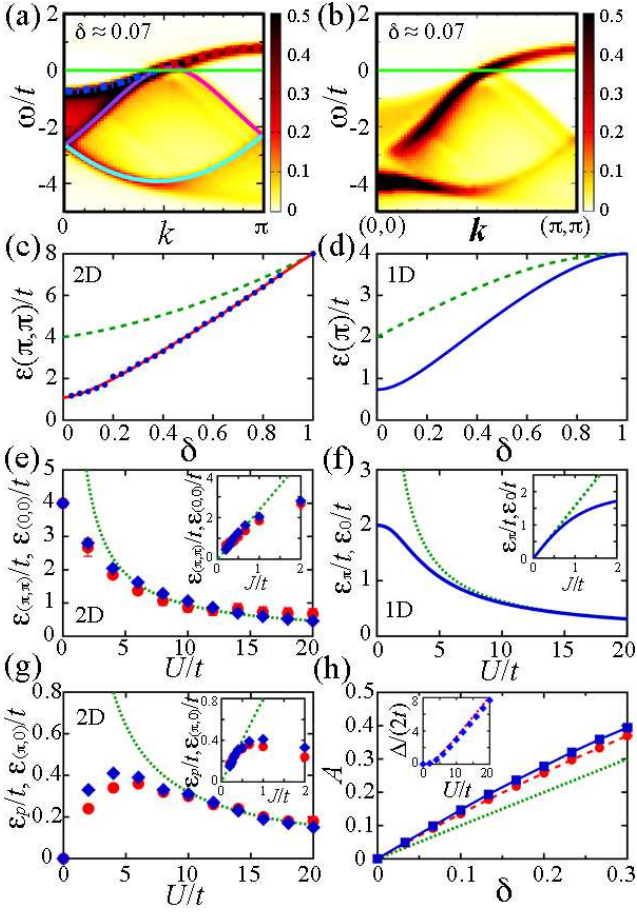}
\caption{(a) $A(k, \omega)t$ in the lower Hubbard band (LHB) for $U/t = 8$ at $\delta \approx 0.07$ 
in the 1D Hubbard model obtained using the dynamical density-matrix renormalization group method \cite{Kohno1DHub}. 
Curves indicate dominant modes obtained using exact solutions \cite{Kohno1DHub}. 
(b) RPA results for $t_{\perp} = t$ obtained using the data in (a). 
In (a,b), straight solid green lines indicate $\omega = 0$. Gaussian broadening is used with standard deviation $\sigma = 0.1t$. 
(c) $\epsilon(\pi,\pi)$ [energy of the mode for $\omega>0$ in the LHB at ($\pi$,$\pi$)] for $U/t = 8$ [Figs. \ref{fig:Fig1}(i--k,p)] (blue circles). 
Solid red curve indicates a fit. Dashed green curve shows results for $U = 0$. 
(d) Same as (c) but for 1D. Solid blue curve shows $\epsilon(\pi)/t$ for $U/t = 8$ \cite{Kohno1DHub}. 
In (d,f), $\epsilon(\pi)$, $\epsilon_{\pi}$, and $\epsilon_0$ are defined in the 1D Hubbard model similarly to $\epsilon(\pi,\pi)$, $\epsilon_{(\pi,\pi)}$, and $\epsilon_{(0,0)}$, respectively. 
(e) $\epsilon_{(\pi,\pi)}/t$ [$\epsilon(\pi,\pi)/t$ extrapolated to $\delta \rightarrow 0$] (red circles) and $\epsilon_{(0,0)}/t$ [Fig. \ref{fig:Fig1}(l)] (blue diamonds). 
Dotted green line indicates  $\sqrt 2 v_{2D}/t$, where spin-wave velocity of the 2D Heisenberg model $v_{2D} \approx 1.18\sqrt 2 J$ \cite{Singh} ($J = 4t^2/U$). 
Inset shows $J/t$ dependence. 
(f) Same as (e) but for 1D. Solid blue curve shows $\epsilon_{\pi}/t (= \epsilon_0/t)$ \cite{Kohno1DHub}. 
Dotted green line indicates $v_{1D}/t$, where spin-wave velocity of the 1D Heisenberg model $v_{1D}=\pi J/2$ \cite{desCloizeaux}. 
(g) $\epsilon_p/t$ (blue diamonds) and $\epsilon_{(\pi,0)}/t$ (red circles) [Fig. \ref{fig:Fig1}(l)]. 
Dotted green line indicates a fit in the large $U/t$ regime, assuming $\epsilon_p$, $\epsilon_{(\pi,0)} \propto J$. Inset shows $J/t$ dependence. 
(h) Spectral weight $A$ for $\omega > 0$ in the LHB. Blue squares with solid line show 2D results for $U/t = 8$. 
Red circles with dashed line denote 1D results for $U/t = 8$ taken from ref. \cite{Kohno1DHub}. Dotted green line indicates results for $t = 0$ \cite{Eskes}. 
Inset shows the Mott gap $\Delta$ between the LHB and the upper Hubbard band at $\delta=0$. Blue diamonds show 2D results. Dotted red curve denotes 1D results \cite{LiebWu}.}
\label{fig:Fig2}
\end{figure}

{\it Properties in the (0,0)--($\pi$,$\pi$) direction.$-$} 
The behavior in the (0,0)--($\pi$,$\pi$) direction [Figs. \ref{fig:Fig1}(i--l)] resembles that of the 1D Hubbard model [Fig. \ref{fig:Fig2}(a)]. 
However, the spectral weight near (0,0) is shifted to lower energies. 
This can be interpreted using the weak-interchain-hopping approximation. 
Up to the first order of interchain hopping $t_{\perp}$, we obtain $G^{-1}({\bm k}, \omega) = G_{1D}^{-1}(k_x, \omega) - t_{\perp}({\bm k})$ \cite{RPAArrigoni,KohnoNP} 
using the 1D Green function $G_{1D}(k_x, \omega)$ and $t_{\perp}({\bm k}) = -2t_{\perp}\cos k_y$. 
In this approximation (called RPA), the spectral weight of a chain 
shifts to higher [lower] energies for $t_{\perp}({\bm k}) > 0$ [$t_{\perp}({\bm k}) < 0$] \cite{KohnoNP,KohnoQ1DH}. 
Thus, the spectral weight near (0,0) [$t_{\perp}({\bm k}) \approx -2t < 0$] shifts to lower energies, 
whereas that near ($\pi/2$,$\pi/2$) [$t_{\perp}({\bm k}) \approx 0$] is almost unaffected by interchain hopping [Fig. \ref{fig:Fig2}(b)]. 
As a result, the dispersion relation exhibits a kink between (0,0) and ($\pi/2$,$\pi/2$) [Figs. \ref{fig:Fig1}(i--l)]. 
This kink can be identified as the high-energy giant kink observed in high-$T_c$ cuprates \cite{Graf,ScalapinoKink}. 
In high-$T_c$ cuprates, a nearly vertical dispersion relation known as a waterfall \cite{Graf} has been observed just below the kink, 
involving smaller spectral weights than those above the kink. 
The RPA is too simple to explain the reduction in spectral weight. 
Nevertheless, the results obtained using CPT [Figs. \ref{fig:Fig1}(i--l,p)], where higher-order $t_{\perp}$-corrections are involved within the clusters, show 
that waterfall behavior appears in the 2D Hubbard model \cite{ScalapinoKink}. 
\par
The RPA argument above allows us to trace the origins of the dominant modes in Figs. \ref{fig:Fig1}(i--l) back to those of 1D \cite{Kohno1DHub,DDMRGAkw,Shadowband}. 
In Figs. \ref{fig:Fig1}(i--k), the mode for $\omega > 0$ originates from the upper edge of the spinon-antiholon continuum [Fig. \ref{fig:Fig2}(a), dashed red curve]. 
The mode for $-1\lesssim\omega/t<0$ and $0 \le k_x \approx k_y \lesssim \pi/2$ primarily originates from the spinon mode [Fig. \ref{fig:Fig2}(a), dashed-dotted blue curve]. 
The mode for $-4 \lesssim \omega/t \lesssim -1$ primarily originates from the holon mode [Fig. \ref{fig:Fig2}(a), solid purple curve]. 
The mode almost extending over the Brillouin zone for $-6 \lesssim \omega/t \lesssim -4$ originates from the shadow band [Fig. \ref{fig:Fig2}(a), solid light blue curve]. 
In Fig. \ref{fig:Fig1}(l), the modes originating from the 1D spinon, holon, and shadow-band modes can be identified similarly. 
The mode originating from the 1D spinon mode and that from the 1D holon mode have been observed in high-$T_c$ cuprates \cite{Graf}. 
The mode originating from the 1D shadow band should also be observed if other bands outside the 2D Hubbard model do not intervene. 
%Later, we will confirm the validity of the above identification in relation to the spin excitation. 
\par
A characteristic feature of the Mott transition is the doping-induced spectral-weight transfer from the upper Hubbard band (UHB) 
to the lower Hubbard band (LHB) \cite{DagottoRMP,KotliarRMP,PhillipsRMP,ImadaCofermion,Eskes,PreussQP,PreussPG,Stanescu,SakaiImada,Kohno1DHub}. 
%However, its nature in 2D remains controversial. 
As shown in Figs. \ref{fig:Fig1}(i--k), the spectral weight transferred to the LHB for $\omega > 0$ is primarily carried by the mode originating from the upper edge of the spinon-antiholon continuum. 
The energy of the mode at ($\pi$,$\pi$) [$\epsilon(\pi,\pi)$] does not reach zero even in the $\delta \rightarrow 0$ limit [Fig. \ref{fig:Fig2}(c)], 
and its spectral weight gradually disappears as $\delta \rightarrow 0$ [Fig. \ref{fig:Fig2}(h)]: 
the mode remains dispersing with the spectral weight fading away as $\delta \rightarrow 0$ [Fig. \ref{fig:Fig1}(p)]. 
This feature is the same as that in 1D [Figs. \ref{fig:Fig2}(d,h)] \cite{Kohno1DHub} 
and contrasts with that of a doped band insulator and with that of a Fermi liquid as the effective mass $m^* \rightarrow \infty$ 
where the effective bandwidth $\epsilon^*({\bm k}) \propto 1/ m^*$~\cite{ImadaRMP}. 
\par
We consider its relationship to the spin-wave mode at $\delta=0$ \cite{AndersonSW}. As shown in Fig. \ref{fig:Fig1}(p), 
the dispersion relations of the mode for $\omega>0$ and the mode originating from the 1D spinon mode for $\omega<0$ in the (0,0)--($\pi$,$\pi$) direction 
behave as $\epsilon_{2D}(k,k)=-\sqrt 2 v_{2D} \cos k$ near the Mott transition in the large $U/t$ regime, 
where $v_{2D}$ denotes the spin-wave velocity of the 2D Heisenberg model \cite{Singh}. 
In fact, as shown in Fig. \ref{fig:Fig2}(e), the bandwidth of the mode for $\omega>0$ in the LHB in the $\delta\rightarrow 0$ limit [$\epsilon_{(\pi,\pi)}$] and 
that of the mode originating from the 1D spinon mode at $\delta=0$ [$\epsilon_{(0,0)}$] [Fig. \ref{fig:Fig1}(l)] 
reasonably agree not only with each other but also with $\sqrt 2 v_{2D}$ in the large $U/t$ regime. 
Thus, the modes lead continuously to the spin-wave mode at $\delta=0$ whose dispersion relation in the (0,0)--($\pi$,$\pi$) direction is expressed 
as $E(k,k)=\left|\epsilon_{2D}(k-\pi/2,k-\pi/2)\right|$ \cite{AndersonSW,Singh}. 
The continuous evolution to the spin-wave mode is consistent with the scaling behavior of spin correlations \cite{ImadaRMP,Furukawa,KohnotJ}. 
In the extremely large $U/t$ regime, the ferromagnetic fluctuation arising from Nagaoka ferromagnetism \cite{Nagaoka} 
could dominate the antiferromagnetic fluctuation near the Mott transition, an outcome that is beyond the scope of the present study. 
Note that the relationship to the spin-wave mode is essentially the same as that in 1D \cite{Kohno1DHub}: 
in 1D, the dispersion relations of the mode for $\omega>0$ in the LHB and the spinon mode 
reduce to $\epsilon_{1D}(k)=-v_{1D} \cos k$ in the $\delta\rightarrow 0$ limit in the large $U/t$ limit [Figs. \ref{fig:Fig2}(a,f)] \cite{Kohno1DHub}, 
where $v_{1D}$ denotes the spin-wave velocity of the 1D Heisenberg model \cite{desCloizeaux}. 
The modes in 1D lead continuously to the spin excitations at $\delta=0$ whose dominant mode shows $E(k)=\left|\epsilon_{1D}(k-\pi/2)\right|$ \cite{desCloizeaux}. 
\par
We next discuss hole pocket behavior. In 1D, the region between the two gapless points at $k=\pi(1-\delta)/2$ and $\pi(1+3\delta)/2$ 
[intersections of solid purple and pink curves with $\omega=0$ in Fig. \ref{fig:Fig2}(a)] 
can be regarded as a hole pocket \cite{Kohno1DHub}. 
Its signature should persist near the Mott transition even in 2D because $t_{\perp}({\bm k}) \approx 0$ near ($\pi/2$,$\pi/2$). 
The small intensity at the intersection of the upper edge of the continuum [bending back near ($\pi/2$,$\pi/2$)] with $\omega = 0$ for $k_x \approx k_y \gtrsim \pi/2$ 
[Figs. \ref{fig:Fig1}(j,k,n--p)] might be regarded as a signature for hole pocket behavior as in 1D. 
\par
{\it Properties near ($\pi$,0).$-$} 
As shown in Figs. \ref{fig:Fig1}(j--l), the dominant mode near ($\pi$,0) near the Mott transition is located below $\omega=0$, whose dispersion relation is anomalously flat. 
This anomalously flat dispersion relation has been found in numerical simulations \cite{PreussPG,Bulut,DagottoFlatband} and in high-$T_c$ cuprates \cite{ShenRMP}. 
Below, we discuss its relationship to the pseudogap and Fermi arc. 
\par
Near the Mott transition, the main peak of $A(\omega) [\equiv \int d{\bm k}A({\bm k}, \omega)/(2\pi)^2]$ is located below $\omega = 0$ [Fig. \ref{fig:Fig1}(k), rightmost panel]. 
Namely, the spectral weight decreases from the peak value as $\omega \rightarrow 0$. This feature is called a pseudogap. 
To discuss the pseudogap quantitatively, we define two energies: 
$\epsilon_p$ as the energy difference between the peak in $A(\omega)$ and the top of the LHB at $\delta=0$ [Fig. \ref{fig:Fig1}(l)], 
and $\epsilon_{(\pi,0)}$ as the energy difference between the flat mode at ($\pi$,0) and the top of the LHB at $\delta=0$ [Fig. \ref{fig:Fig1}(l)]. 
Figure \ref{fig:Fig2}(g) shows that the pseudogap defined in $A(\omega)$ is primarily due to the flat mode at ($\pi$,0) 
and that the pseudogap is induced by interactions even without further-neighbor hopping. 
In addition, this figure indicates that the pseudogap is proportional to $J$ ($= 4t^2/U$) [$\epsilon_p \approx \epsilon_{(\pi,0)} \approx 0.8J$] in the large $U/t$ regime, 
implying that it will be related to the antiferromagnetic fluctuation \cite{PreussPG,Bulut}. 
The pseudogaps defined by $A(\omega)$ and by the flat mode decrease as $\delta$ increases [Figs. \ref{fig:Fig1}(j,k)], 
because the chemical potential is lowered with the flat mode almost unchanged. 
The pseudogap closes at a $\delta$ value where the peak in $A(\omega)$ or the flat mode crosses $\omega = 0$. 
\par
The flat mode also causes Fermi arc behavior in the pseudogap regime. In the (0,0)--($\pi$,0) direction, there is no dominant mode crossing $\omega = 0$ in the pseudogap regime, 
because the flat mode is located below $\omega = 0$ [Figs. \ref{fig:Fig1}(j,k)]. 
Thus, there is no Fermi surface along this direction \cite{PreussPG,Bulut,Stanescu,SakaiImada}. 
This feature can also be confirmed from $n({\bm k}) [\equiv \int_{-\infty}^0d\omega A({\bm k}, \omega)]$ [Figs. \ref{fig:Fig1}(f--h), solid blue curves with hatches]: 
$n({\bm k})$ along this direction in the pseudogap regime [Figs. \ref{fig:Fig1}(f,g), rightmost panels] almost saturates to that at $\delta=0$ [Fig. \ref{fig:Fig1}(h), rightmost panel]. 
For the ($\pi$,0)--($\pi$,$\pi$) direction, two possibilities have been discussed: the absence of the mode crossing $\omega = 0$, 
which leads to hole pockets \cite{PreussPG,Stanescu,SakaiImada}, and the presence of the mode forming the large Fermi surface centered around ($\pi$,$\pi$) \cite{Bulut}. 
Figures \ref{fig:Fig1}(k,o) show intermediate behavior. 
Namely, the mode crossing $\omega = 0$ in the ($\pi$,0)--($\pi$,$\pi$) direction can be identified as that originating from the upper edge of the spinon-antiholon continuum, 
whose spectral weight fades away as $\delta\rightarrow 0$. As a result, the spectral weight almost disappears along this direction around $\omega = 0$. 
This leads to the behavior that can be regarded as a Fermi arc (disconnected portion of the large Fermi surface) \cite{ShenRMP} [Fig. \ref{fig:Fig1}(o)]. 
This behavior contrasts with that of the large Fermi surface in the large doping regime [Fig. \ref{fig:Fig1}(m)]. 
Fully quantitative explanation for the Fermi arc observed in high-$T_c$ cuprates requires further study. 
Also, properties in a very small energy scale inaccessible with ($4\times 4$)-cluster CPT are beyond the scope of this Letter. 
\par
{\it Mott gap.$-$} 
%We consider the physics of the Mott gap. 
In a (doped) Mott insulator, the Mott gap is due to Coulomb interactions, 
whereas the band gap in a (doped) band insulator is due to a chemical potential difference between sites or orbitals within a unit cell. 
A deeper insight into Mott physics can be obtained by tracing the origin back to 1D. 
In 1D, the quasiparticle responsible for the Mott gap has been identified as that defined by the $k$-$\Lambda$ string in the Bethe ansatz and called the doublon \cite{Kohno1DHub}, 
which can be regarded as a pair of electrons \cite{Takahashi1DHub}. 
By the presence of the doublon, the UHB can be distinguished from the LHB \cite{Kohno1DHub}. 
The doublon is not a double occupancy because the latter exist in the LHB as well as the UHB for $U/t < \infty$. 
Because the Mott gap in 2D can change into that in 1D by reducing the interchain hopping, 
the quasiparticle that determines the energy scale of the UHB in 2D (doublon in 2D) should be a descendant of the 1D doublon, i.e., a pair of electrons rather than a double occupancy. 
The similarities in the spectral weight transferred by doping and those in the Mott gap at $\delta=0$ between 1D and 2D  [Fig. \ref{fig:Fig2}(h)] seem to support the traceability. 
Apart from the presence of the doublon, we can identify the dominant modes in the UHB [Figs. \ref{fig:Fig1}(a--d)] as in the LHB for $\omega < 0$, 
noting the particle-hole symmetry at $\delta = 0$, as in 1D \cite{Kohno1DHub}.
\par
Noting that the high-energy magnetic excitations in a magnetic field in quasi-1D Heisenberg antiferromagnets, 
which can be regarded as repulsively interacting hard-core boson systems, have been explained using two-string solutions \cite{KohnoQ1DH,Kohno1DH,KohnoHighE}, 
we can generalize the concept of Mott physics: 
the generalized doublon (a pair of single particles definable by a string with a length of two in 1D \cite{Kohno1DHub,Kohno1DH,KohnoHighE} or 
its descendant in higher dimensions \cite{KohnoQ1DH}) induced by repulsive interactions will be responsible for the high-energy states 
without multi-site (multi-orbital) unit cells, in contrast to band theory. 
The Mott physics could also be explored in cold atomic systems. 
\par
{\it Summary.$-$} 
Anomalous spectral features observed in high-$T_c$ cuprates, such as the pseudogap, Fermi arc, flat band, doping-induced states, spinon-like and holon-like branches, 
as well as kink and waterfall in the dispersion relation, were explained in a unified manner as properties of the 2D Hubbard model near the Mott transition. 
The properties near ($\pi$,0) are characterized by the flat mode unlike 1D features. 
The physics of the Mott gap was examined by tracing the origin back to 1D. 
The Mott transition is characterized by a dispersing mode that leads continuously to the spin-wave mode of the Mott insulator 
with the spectral weight fading away toward the Mott transition due to charge freezing. 
The loss of charge character from the dispersing mode will be a general feature of Mott transitions, 
which contrasts with the transition to a band insulator without spin-charge separation. 
\par
This work was supported by KAKENHI 22014015 and 23540428, and WPI for Materials Nanoarchitectonics.
%\bibliography{apssamp}

\end{document}